\def\spose#1{\hbox to 0pt{#1\hss}}
\def\approxlt{\mathrel{\spose{\lower 3pt\hbox{$\sim$}}
        \raise 2.0pt\hbox{$<$}}}
\def\approxgt{\mathrel{\spose{\lower 3pt\hbox{$\sim$}}
        \raise 2.0pt\hbox{$>$}}}
\def\approxpropto{\mathrel{\spose{\lower 3pt\hbox{$\sim$}}
        \raise 2.0pt\hbox{$\propto$}}}
\mathchardef\twiddle="2218

\def\multleft#1{\hbox to size{\vbox {\halign {\lft{##}\cr #1}}\hfill}\par}
\def\multright#1{\hbox to size{\vbox {\halign {\rt{##}\cr #1}}\hfill}\par}
\def\Mdot{\hbox{$\dot M$}}
\def\mdot{\hbox{$\dot m$}}
\def\<{\thinspace}

\def\cm{{\rm\thinspace cm}}
\def\erg{{\rm\thinspace erg}}

\def\g{{\rm\thinspace g}}

\def\keV{{\rm\thinspace keV}}

\def\km{{\rm\thinspace km}}
\def\kpc{{\rm\thinspace kpc}}

\def\Mpc{{\rm\thinspace Mpc}}
\def\Msun{\hbox{$\rm\thinspace M_{\odot}$}}

\def\s{{\rm\thinspace s}}
\def\yr{{\rm\thinspace yr}}

\def\ergps{\hbox{$\erg\s^{-1}\,$}}

\def\gps{\hbox{$\g\s^{-1}\,$}}
\def\kmps{\hbox{$\km\s^{-1}\,$}}

\def\Msunpyr{\hbox{$\Msun\yr^{-1}\,$}}
\def\pcm{\hbox{$\cm^{-3}\,$}}

\def\psqcm{\hbox{$\cm^{-2}\,$}}

\def\kmpspMpc{\hbox{$\kmps\Mpc^{-1}$}}

\documentclass[]{article}
\usepackage{tdm_emulateapj}
\usepackage{psfig}
\newcommand\beq{\begin{equation}}
\newcommand\eeq{\end{equation}}

\begin{document}

 \title{ACCRETION ONTO NEARBY SUPERMASSIVE BLACK HOLES: {\it Chandra}
  CONSTRAINTS ON THE DOMINANT CLUSTER GALAXY NGC 6166}

\author{Tiziana~Di~Matteo\footnote{{\em Chandra} Fellow}}
\affil{Harvard-Smithsonian Center for Astrophysics, 60 Garden St., Cambridge, MA 02138; 
tdimatteo@cfa.harvard.edu}
\author{Roderick M. Johnstone, Steven W. Allen and Andrew C. Fabian}
\affil{Institute of Astronomy, Madingley Road, Cambridge, CB3 OHA, UK; 
rmj,swa,acf@ast.cam.ac.uk }

\affil{}

\begin{abstract}

{\it Chandra} observations of low-luminosity supermassive black holes
in nearby elliptical galaxies provide tight limits on both their
nuclear luminosities and on their Bondi accretion rates. We examine
{\it Chandra} constraints on NGC 6166, the dominant galaxy in the
cluster Abell 2199, which hosts a $\sim 10^9 \Msun$ black hole. We
measure a nuclear X-ray luminosity $L_{x,1\keV} \sim 10^{40} \ergps$
and show that the density and temperature profiles of the hot
interstellar medium imply a Bondi accretion rate 
of $\Mdot_{\rm Bondi} \approxlt 3 \times 10^{-2} \Msunpyr$. This
accretion rate predicts a nuclear luminosity of $\sim 10^{44}
\ergps$ for a canonical radiative efficiency of 10\%.
Unless the Bondi estimate is inappropriate and/or the accretion rate
onto the black hole is significantly reduced, the observed nuclear flux
constrains the radiative efficiency of the accretion flow to be $\eta
\approx 10^{-5}$. We show that low-radiative efficiency accretion flows 
with radial density profiles $\rho \approxpropto r^{-3/2}$ (and not
significantly flatter) can explain the observed nuclear X-ray
luminosity but that the power output from the jets in NGC 6166 is also
important to the energetics of the system.

\end{abstract}

\section{Introduction}

Recent dynamical measurements of black hole masses in nearby galaxies
have transformed black hole studies, allowing us to quantify directly
their expected accretion power. In those cases in which the fuel
supply to the black holes can be estimated, it has become apparent
that the expected power far exceeds the observed luminosity.  This
situation is perhaps best illustrated by the case of nearby elliptical
galaxy nuclei (e.g.Fabian \& Rees 1995; Di Matteo et al. 2000a,
hereafter DM00a and references therein) and our Galactic center
(e.g.~Narayan et al. 1998 and references therein). There are only two
possible explanations for the low-luminosities of nearby black holes:
(a) the accretion occurs at extremely low rates or (b) the accretion
occurs at low radiative efficiencies as predicted, for example, by
advection dominated accretion models (e.g. Rees et al. 1982; Narayan
\& Yi 1994,1995; Abramowicz et al.~1995).

Assuming that the accretion in elliptical galaxies occurs primarily
from the hot, quasi-spherical interstellar medium (ISM), Bondi theory
(1952) can be used to estimate the accretion rates onto the
supermassive black holes. Such estimates require accurate measurements
of both the density and temperature of the hot ISM at the 'Bondi
accretion radius', the radius where the gravitational potential of
the central black holes begins to dominate the dynamics of the hot
gas.

In order to unambiguously determine whether the low luminosities of
nearby black holes are due to a low radiative efficiency in the
accreting gas, it is also necessary to measure the nuclear power. When
combined with the estimated accretion rates, this gives us a direct
measurement of the radiative efficiency of accretion, $\eta$.

Thanks to its high spatial resolution and sensitivity, the {\it
Chandra} X-ray observatory is able, for the first time, to detect
nuclear X-ray point sources in nearby galaxies and provide us with
direct measurements of their luminosities.  Chandra also allows us to
measure the central densities and temperatures of the ISM close to the
accretion radii of the central black holes, and therefore determine
the Bondi accretion rates in these systems to much greater accuracy
than before.

In this {\it Letter} we explore the implications of Chandra
observations of the giant elliptical galaxy NGC 6166. We report the
detection of a nuclear point source in the galaxy with a luminosity
$L_{x} \sim $ a few $10^{40}\ergps$. We measure the central X-ray gas
density and temperature and calculate the Bondi accretion rate. We
show that if the central black hole of $\sim 10^9
\Msun$ is fed at the estimated Bondi rate,
the inferred accretion radiative efficiency is $
\eta \approxlt 10^{-5}$. 
We show that at the given accretion rate ADAF models can explain the
observed nuclear luminosity. The presence of outflows in the accretion
flows is also consistent with the present constraints. We discuss
possible reasons why the Bondi accretion rate estimate might be
inappropriate and show that a significant contribution to the nuclear
flux is likely to be due to emission from the jets.

\section{Nuclear luminosity and accretion rate}

NGC 6166 ($z=0.031$) is a cD galaxy at the center of the cooling flow
cluster A2199, and is therefore located in a highly gas-rich
environment. The central black hole mass is estimated by Ferrarese \&
Merritt (2000) using the correlation between the masses of black holes
and the velocity dispersions of their host bulges ($M_{BH}
\propto \sigma_c^{4.7}$).
For NGC 6166, $\sigma_{c} = 326$\kmps~ which gives $ M_{BH} \sim 10^{9}
\Msun$, implying a discrepancy of about an order of magnitude with the
value obtained by Magorrian et al. (1998), who estimated $M_{BH} \sim
10^{10} \Msun$.

NGC 6166 is known to host an active nucleus, classified as an FR I
source, which powers two symmetric parsec-scale radio jets and radio
lobes. The jets are close to the plane of the sky (Giovannini et
al. 1998). Beaming should not therefore be important. Given the
measured black hole mass, the Eddington luminosity of NGC 6166 is
$L_{\rm Edd} \sim 10^{47}
\ergps$. However, the observed bolometric luminosity is $\approxlt$ $10^{41} \ergps$. 

We adopt the cosmological parameters $H_0 =65 \kmpspMpc$, $\Omega = 1$
and $\Lambda = 0$, which imply a distance to NGC 6166 of $142 \Mpc$).

\subsection{$L_{x}$}

The {\it Chandra} observations of the A2199 cluster were carried out
on 1999 December 11 using the Advanced CCD Imaging Spectrometer and 
back-illuminated CCD detectors. The net exposure time was 14.95 ks. 
Here, we only highlight results for the central, dominant
galaxy NGC 6166; details of the cluster and point source analysis are
discussed by Johnstone et al. (2000).

The Chandra data for NGC 6166 reveal the presence of a nuclear
point-like X-ray source (detected over background at 7 sigma
significance) with a surface brightness distribution consistent with
the point-spread function (PSF) of the instrument ($\sigma \sim 0.45$
arcsec). The X-ray source is coincident with the peak of the emission
from the central radio source (Giovannini et al.~1998).  In order to
accurately determine the central point source flux, we have modeled
the central surface brightness profile using a simple power-law
($I_{\rm X} \propto r^{-\alpha}$; accounting for the underlying ISM)
plus Gaussian model, convolved with the instrumental PSF. We find that
the surface brightness profile within the innermost 10 kpc radius
region can be well-described by a power-law model of slope $\alpha =
0.4\pm 0.025$ and a point source with $55\pm 27$ counts ($1\sigma$
errors; where the error in the counts is dominated by the uncertainty
in the slope of the density profile).

We have extracted a spectrum for the central source using a region of
$\sim$~1.5 arcsec radius (which should contain $\approxgt 90$ per cent
of the flux from the source) and used a background spectrum taken from
the $1.5-2.5$ arcsec annulus. Given the faintness of the central
source and the moderate exposure time of the observation, the spectral
parameters are not firmly constrained and the data can be described by
either a power law or a thermal plasma model (mekal). The power-law
model gives a photon index $\Gamma = 1.54^{+0.46}_{-0.45}$ assuming
Galactic absorption, and a flux density at 1
\keV of $7 \pm 2 \times 10^{-15}\erg\cm^{-2}\s^{-1}\keV^{-1}$, where all
parameters are quoted at a $1\sigma$ confidence.  (Allowing for
intrinsic absorption, the maximum intrinsic column density is $N_{H}
\sim 2 \times 10^{21}$ atom$\psqcm$ for $\Gamma = 2.9$ and $F(1 \keV)
= 14 \times 10^{-15}\erg\cm^{-2}\s^{-1}\keV^{-1}$; see Fig.~3 and
contour plots in Johnstone et al.~2000). The flux determined from the
spectrum is consistent with the value measured from the surface
brightness profile, assuming a consistent spectral model for the
nuclear emission.

\subsection{$\Mdot$}

We can estimate the accretion rate from the hot ISM onto the central black
hole using Bondi accretion theory. Matter passing within the
accretion radius $R_{\rm A} \approx (c/c_s)^2 R_S$ is assumed to be
accreted. Here $c$ is the speed of light, $c_{\rm s}
\sim 10^4T^{1/2}$ cm s$^{-1}$ is the sound speed (with $T$ the gas
temperature), and $R_{\rm S} = 2GM/c^2$ the Schwarzschild radius of
the black hole.

We have measured the temperature and density profile of the
intracluster gas within the central 100 \kpc~radius region. Direct
measurements of these quantities are only possible down to within $r
\sim 1$ arcsec of the nucleus, due to contamination from the nuclear
point source.  The filled triangles in Fig.~1 show the temperature
profile determined by fitting the annular counts with a single
temperature thermal emission model (mekal) acted on by Galactic and
(free-fitting) intrinsic absorption. The open squares show the
deprojected temperature profile in which the inner regions are
corrected for the emission from further out seen in projection. The
temperature is higher in the outer regions, where the cluster
potential dominates and decreases (and flattens off) in the inner
regions where the galaxy potential takes over.

The density profile shown in Fig.~2 was derived by deprojecting the
surface brightness profile (e.g.~Fabian et al. 1981) and matching the
deprojected temperature profile. The density profile (Fig.~2) is
steeper in the outer regions ($r \approxgt 10 \kpc$) and follows the
canonical cooling-flow profile of $\rho \approxpropto r^{-1}$. Inside
this radius the density profile flattens to a value $\rho
\propto r^{-0.45 \pm 0.09}$ (we exclude the central 1.5 arcsec region, 
which is affected by the point source).

We take $kT\sim 1.3 \keV$ as measured in the innermost regions.  This
gives \beq R_A \sim 0.06 T^{-1}_{1.3} M_9 \;\;\; \kpc, \eeq where $M_9
= 10^9 \Msun$ which corresponds to to $\sim 0.08 $ arcsec.

The accretion rate is related to the density and temperature by the
continuity equation, with the velocity $v =c_{s}$;
\begin{equation} 
\Mdot = 4\pi R_{\rm A}^2\rho_{\rm A}c_{\rm s}(R_{\rm A})
\end{equation}
where $\rho_{\rm A}$ and $c_{\rm s}(R_{\rm A})$ are the density and
sound speed at the accretion radius.

Given that the density and temperature cannot be measured directly at
the accretion radius we assume $kT =1.3^{+0.6}_{-0.7} \keV$ and
estimate the density at $R_A$ by extrapolating the best-fit power-law
model (see above) which gives $n(R_{\rm A}) \sim 0.6 \pm 0.1 \pcm$.

This implies
\begin{eqnarray} \Mdot_{\rm Bondi}& = &1.3^{+1.2}_{-0.7} \times 10^{24} \;
M_9^2\;\;T_{1.3}^{-3/2} \;n_{0.6} \;\;\;\;\gps \\
&&\sim 3 \times 10^{-2}\;\; \Msunpyr ,\nonumber \end{eqnarray} 

(where we have neglected the uncertainties on $M_{\rm BH}$). Eqn.~(2)
gives an estimate of the accretion at the outer edge of the accretion
flow. As discussed in the next session, the mass accretion rate onto
the black hole may be smaller if e.g.~it decreases with radius because
of an outflow.

This Bondi accretion rate implies $L_{Bondi}=\eta \Mdot_{\rm Bondi} c^2
\sim 10^{44} \ergps$ if $\eta =0.1$, as in a standard radiatively
efficient thin disk. This is about four orders of magnitude greater
than the observed luminosity of the central X-ray source, implying
that the radiative efficiency has to be $\eta
\sim 10^{-5}$ suggesting that hot accretion flows with low radiative
efficiencies may be relevant.

\section{Models}

In a hot accretion flow around a supermassive black hole, the majority
of the emission arises in the radio and X--ray bands. In the radio
band the emission results from synchrotron radiation from the inner
parts of the accretion flow. The X-ray emission is due to either
bremsstrahlung emission or inverse Compton scattering of the soft
synchrotron photons (e.g.~Narayan, Barret \& McClintock 1998)

We measure radii in the flow in Schwarzschild units: $R = rR_{\rm S}$,
where $R_{\rm S} = 2GM_{\rm BH}/c^2$.  We measure black hole masses in
solar units and accretion rates in Eddington units: $M_{\rm BH} = m
\Msun$ and $\Mdot= \mdot \Mdot_{\rm Edd}$. We take $\Mdot_{\rm Edd} =
10L_{\rm Edd}/c^2 = 2.2 \times 10^{-8} m \Msun$ yr$^{-1}$, i.e., with
a canonical 10\% efficiency. We take $r=10^4$ to be the outer radius
of the flow. The Bondi accretion rate in Eddington units is
$\mdot_{\rm Bondi} = 1.3 \times 10^{-3}$.

The predicted spectrum from an ADAF depends (weakly) on the ratio of
gas to magnetic pressure $\beta$, the viscosity parameter $\alpha$,
and the fraction of the turbulent energy in the plasma which heats the
electrons, $\delta$. Here, we fix $\alpha = 0.1$, $\beta = 10$, and
$\delta = 0.1$. The two major parameters, though, are the accretion
rate $\Mdot$ and the black hole mass $M_{BH}$, both of which are
constrained. With $M_{BH}$ assumed, we
normalize the models to the observed {\it Chandra} flux. This gives us
the $\mdot$ required by the models to explain the X-ray emission.
This has to be consistent with our estimate of $\mdot_{\rm Bondi}$.

The solid line in Fig.~3 shows the predicted spectrum for a pure
inflow ADAF model fitted to the 1 keV flux. In this model $\mdot \sim
10^{-3}$, consistent with the Bondi estimate. Comptonization of the
synchrotron emission dominates the X-ray emission in this model. ADAFs
models can therefore explain the observed X-ray emission. At the given
Bondi rate the radiative efficiency of an ADAF is $\sim 10^{-5}$
corresponding to that derived in \S 2.2. Note, however, that in these
models the synchrotron component cannot explain the VLBI radio
emission (Giovannini et al. 1998; Fig.~3). This implies that the
majority of the contribution in the radio band is likely to be due to
emission from the jets (e.g., dotted line).

Following the proposal of Blandford \& Begelman (1999) that mass loss
may be important in hot accretion flows, Di Matteo et al.~(1999,
hereafter DM99; DM00a) and Quataert \& Narayan (1999; QN99) have
constructed models in which a significant fraction of the accreted
mass in an ADAF is lost to an outflow rather than being accreted onto
the central object. In that work (see also Di Matteo, Fabian \&
Carilli 2000b, hereafter DM00b) it was shown that such models are
favored (or required) over the pure inflow ADAF for explaining the
observed spectral energy distributions of elliptical galaxy nuclei.

The importance of outflows in ADAFs can be parameterized by using
different radial density profiles in the flows with $\rho \propto
r^{-3/2 +p}$ and $0 <p \le 1$ (i.e. flatter than in the pure Bondi
inflow with $\rho \propto r^{-3/2}$). This implies mass inflow rates
which satisfy the relation $\Mdot = \Mdot_{\rm out} (r/r_{\rm
out})^p$, where we should satisfy $\Mdot_{\rm out} \sim \Mdot_{\rm Bondi}$ and
$r_{\rm out} = 10^4$ as before. Bremsstrahlung emission in these
models dominated the X-ray emission. This produces a harder X-ray
spectrum, most of which (in the 0.5 - 10 \keV~ band) is produced in the
outer regions of the flow (the temperatures in the inner regions
approach 100 \keV~to 1 MeV) at $r \sim r_{\rm out}$ where $\mdot_{\rm out}
\sim \mdot$  (see details in DM00a; QN99). 
The dashed line shows a model with $p=0.1$ which implies $\mdot_{\rm
out} = 3\times 10^{-3}$. If $p \approxgt 0.1$, $\mdot_{\rm out}$
becomes inconsistent with the Bondi estimate (Eq.3) unless the errors
on $M_{BH}$ are sufficiently large (e.g., for $p=0.2$, $\mdot_{\rm
out} \sim 6 \times 10^{-3}$, so $M_{BH}$ would need to have been
underestimated by a factor $\approxgt 50 \%$).

For certain regimes (e.g.~for $\alpha \approxlt 0.1$) convection might
also become important in hot accretion flows, (e.g.~Stone, Pringle \&
Begelman 1999; Quataert \& Gruzinov 2000; Narayan, Igumenshchev \&
Abramowicz 2000). Although CDAFs have a very different structure than
ADAFs, they have density profiles $\rho \propto r^{-1/2}$
(corresponding to the $p=1$ case of the outflow models above). Ball,
Narayan \& Quataert (2000) have constructed spectral models for CDAFs
similar to those considered by DM99 and QN99 for
outflows\footnote{Note that in a CDAF the Bondi accretion rate
estimate is not appropriate as convection and possibly conduction
significantly alter the rate at which matter is fed into the flow. The
accretion rates (at large radii) are expected to be lower than in the
ADAFs}.  Taking $p=1$, even with $\mdot_{\rm out} \sim \mdot_{\rm
Bondi}$, implies very small contributions from a CDAF to the X-ray
flux.  A CDAF if present cannot explain the observed emission.

Finally, the dotted line in Fig.~3 shows an example of non-thermal jet
models for the full band spectrum. It is clear that the observed radio
jets must contribute significantly to the observed radio and optical
emission even if an ADAF (with or without outflows) is also
present. Fig.~3 shows that the full spectral energy distribution of
NGC 6166 can be easily explained by the emission from different
compact regions emitting self-absorbed non-thermal synchrotron
radiation. The typical parameters for these components are $B \sim 7
\times 10^{-4} \nu_{t
\rm {GHz}} T_{B 12}^{-2}$ G and $\tau_T \sim 0.4 T_{B 12}^{5} \nu_{t
{\rm GHz}}$; where $T_{B 12} = 10^{12}$ K is the brightness
temperature and $\nu_{t}$ the self-absorption frequency (see Di
Matteo, Carilli \& Fabian 2000 for details, hereafter DM00b). Emission
from a jet is far more efficient even if it originates from relatively
low density regions and can easily give rise to the observed X-ray
flux (see also e.g.~Falcke 1999). Explaining most of the observed
emission as being due to the jets would be consistent with the
unifying model in which FR I radio galaxies are mis-oriented BL Lac
objects and the fact that the radio-to-optical and optical-to-X-ray
spectral indexes of NGC 6166 are consistent with those of radio
selected BL Lac objects (see Fig.~3 in Hardcastle \& Worrall 1999).
However, this interpretation would still not relax the requirement for
a low-radiative efficiency in the accretion flow.

\section{Discussion}
We have examined {\it Chandra} constraints on the low luminosity
active nucleus in NGC 6166. By adopting $M_{BH} =10^{9} \Msun$ and
inferring the ISM density and temperature at $R_{\rm A}$ we
have estimated the Bondi accretion rate to be $\sim 3 \times 10^{-2}
\Msunpyr$. This value is $\approxlt 10$ times lower than the typical
Bondi rates determined in previous studies based on ROSAT High Resolution
Imager (HRI) observations of other nearby elliptical nuclei (e.g., M87,
NGC4649, NGC4472; DM00a) and is more consistent with the upper limits
inferred from high frequency radio studies of the accretion rates at
small radii in the ellipticals (DM99;DM00b; Wrobel \& Herrnstein 2000).

The {\it Chandra} observations of NGC 6166 indicate a nuclear
luminosity of $L_{x, 0.5-7 \keV} \sim 4 \times 10^{40} \ergps$. This
implies that if matter is accreted from the ISM at the Bondi rate the
radiative efficiency of the accretion flow has to be $\eta
\lesssim 10^{-5}$. We have shown that this is consistent with the 
predictions of pure inflow ADAF models. Hot accretion flow models with
significantly flatter density profiles (with strong outflows or
CDAFs), which are favored on theoretical grounds and from radio
observations (DM99,DM00b), cannot explain the observed nuclear flux
(unless $M_{BH}$ is significantly larger). This may imply that most of
the observed flux is dominated by non-thermal emission from the jets.

Although with {\it Chandra} observations we can infer Bondi accretion
rates with much greater confidence than before, for NGC 6166 we can
only measure the ISM densities and temperatures directly at a radius
$\sim 1\kpc$ which is still a factor $\sim 15$ larger than $R_{\rm A}$
(\S 2.2). Given the distance to NGC 6166, previous ROSAT HRI
observations would have only been able to measure the density profile
down to $r\approxgt 10 \kpc $, where we observe the density profile to
steepen. Extrapolation of this steeper density profile to the
accretion radius would then have lead to higher inferred accretion
rates. Even with {\it Chandra}, however, we can still only infer the
ISM density at $R_{\rm A}$ in NGC 6166, which implies that we could
still be overestimating the Bondi accretion rate (if e.g. the density
profile flattened further towards $R_{\rm A}$). Given that in order to
explain the observed nuclear flux, an ADAF requires accretion rates
similar to those currently estimated, a decrease in the Bondi estimate
would make the presence of an ADAF undetectable and their emission
negligible. Similar arguments apply if the temperature closer to
$R_{\rm A}$ increased.

It is also possible that the Bondi calculation itself may be
inappropriate. If strong convection were important, as in CDAFs, the
rates at which matter is fed through the accretion radius could be
significantly reduced. Heating of the ISM by the jets (DM00b) could
also decrease significantly the Bondi accretion radius and hence the
rate at which mass is fed into the accretion flows. If the accretion
rate is low enough, the requirement for low radiative efficiency in
the accretion flow may be significantly relaxed.

We have shown that emission from the jets in these low-luminosity
nuclei provides an important contribution to the overall energy
output of NGC 6166. Determining the variability properties of the
X-ray emission may be important for resolving the jet contribution
from that of an ADAF. Chandra observations of other, closer,
low-luminosity black holes (e.g.,~M87) are needed to further
assess the question of low radiative-efficiency versus low-$\Mdot$
accretion in nearby black holes.

\acknowledgements
We thank Ramesh Narayan for many useful comments. T.\,D.\,M.\ acknowledges
support for this work provided by NASA through Chandra Postdoctoral
Fellowship grant number PF8-10005 awarded by the Chandra Science
Center, which is operated by the Smithsonian Astrophysical Observatory
for NASA under contract NAS8-39073. SWA and ACF thank the Royal
Society for support.

\begin{figure}[t]
\centerline{\psfig{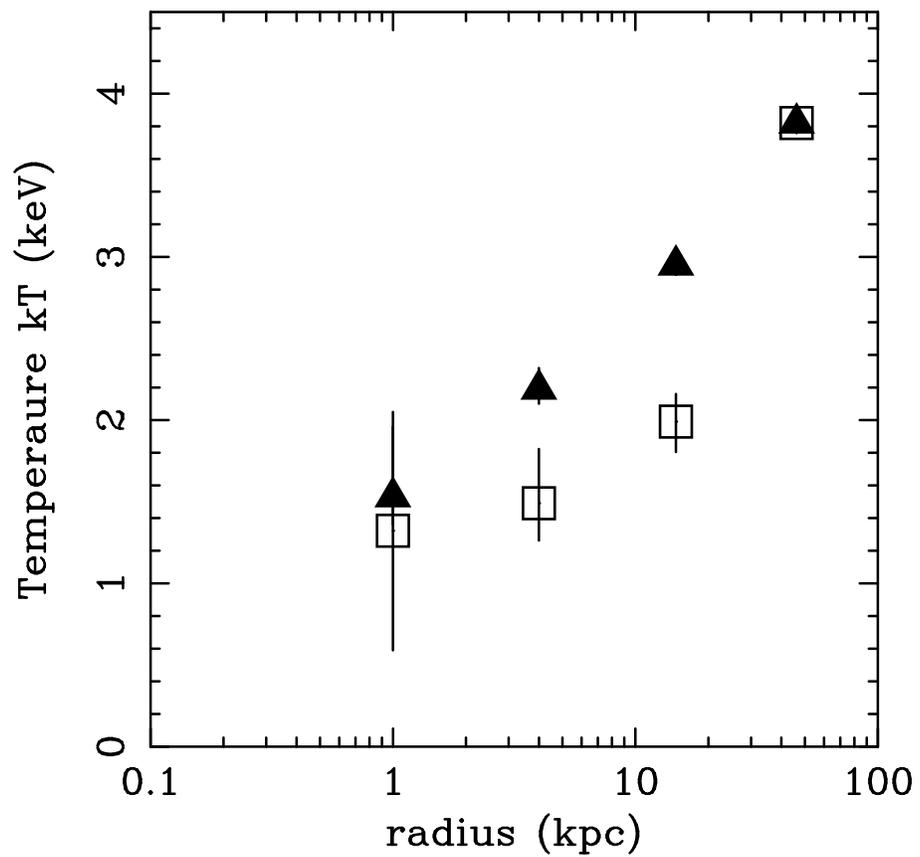}}
\caption{The temperature profile in NGC 6166 measured by {\it Chandra}. 
The open squares show the deprojected temperature measurements
and the filled triangles the observed (projected) values.}
\end{figure}

\begin{figure}[t]
\centerline{\psfig{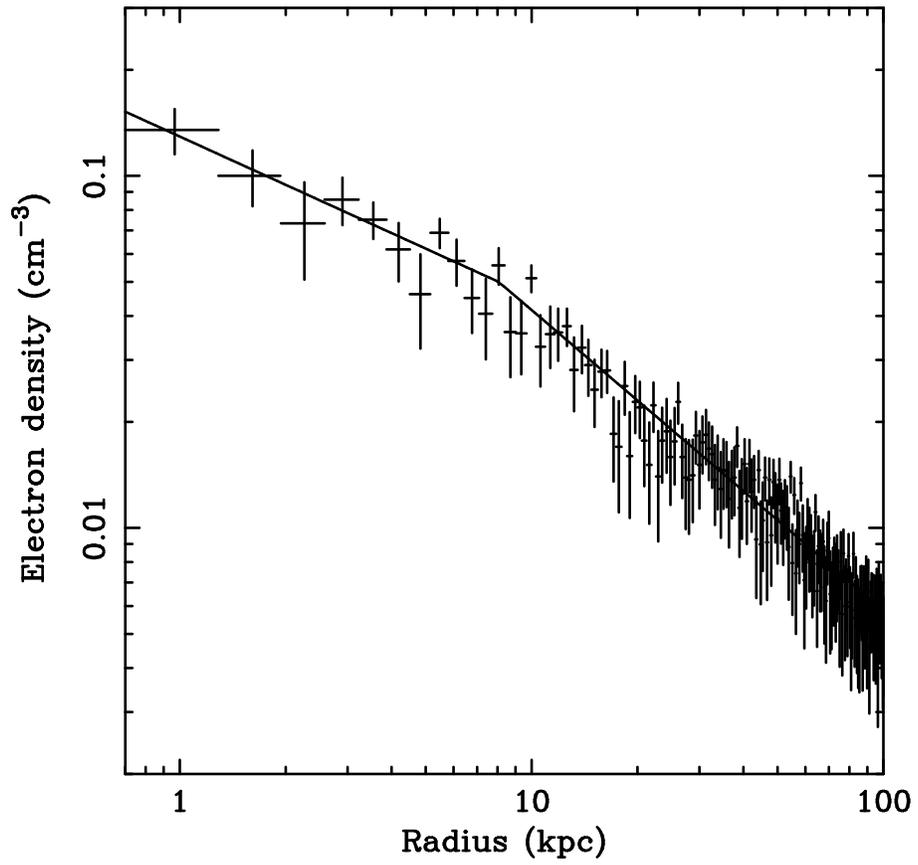}}
\caption{The deprojected density profile in the inner regions of A 2199.}
\end{figure}

\begin{figure}
\centerline{\epsfysize=5.7in\epsffile{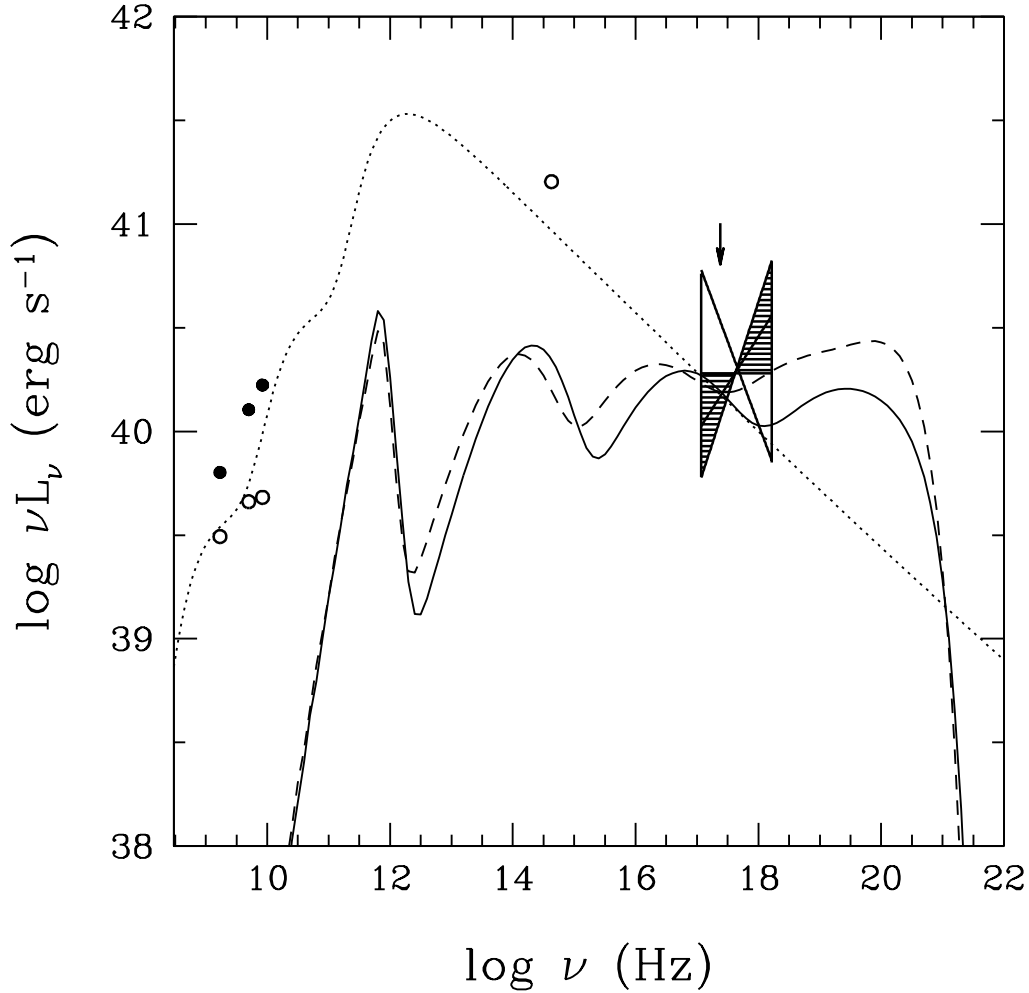}}
\caption{Spectral models calculated for hot accretion flows fitted to the {\it
Chandra} X-ray flux. The solid line is an ADAF model. The required
accretion rate, $\mdot = 8 \times 10^{-4}$, is consistent with the
inferred Bondi value. The dashed line shows a model with
outflows. Here $p=0.1$, $\mdot_{\rm out} = 3 \times 10^{-3}$.  The
dotted line is the sum of self-absorbed, non-thermal synchrotron
emitting regions in the jets. Beaming is ignored since the jets lie
very close to the plane of the Sky (Giovannini et al.~1998). The
solid dots are VLBI flux measurements from Giovannini et al. (1998),
the open circles the VLBI core peak fluxes and the HST (Chiaberghe et
al.~1999) continuum flux and ROSAT HRI upper limit (Hardcastle \&
Worrall 1999). The bow-tie shows the Chandra flux (in the $0.5-7$
\keV~band) for the range of power-law slopes with the hatched area
corresponding to the ones acceptable with Galactic absorption fixed.}

\end{figure}

\end{document}